\documentclass[twocolumn,aps,pra,showpacs,showkeys,superscriptaddress]{revtex4}
\usepackage{amsmath}
\usepackage{graphicx}

\begin{document}

\title{Temporal Behavior of Rabi Oscillation in Nanomechanical QED System
with a Nonlinear Resonator}

\author{X. Xiao}

\affiliation{Department of Physics, Renmin University of China, Beijing, 100872,
China}

\author{M. Y. Zhao}

\affiliation{College of Applied Sciences, Beijing University of Technology, Beijing,
100124, China}

\author{S. M. Yu}

\affiliation{College of Applied Sciences, Beijing University of Technology, Beijing,
100124, China}

\author{Y. B. Gao}

\email[E-mail:]{ybgao@bjut.edu.cn}

\affiliation{College of Applied Sciences, Beijing University of Technology, Beijing,
100124, China}
\begin{abstract}
In nanomechanical QED system, consisting of a charge qubit and a nanomechanical resonator with intrinsic nonlinearity, we study the temporal behavior
of Rabi oscillation in the nonlinear Jaynes-Cummings model. Using
microscopic master equation approach, we solve time evolution of the
density operator describing this model. Also, the probability of excited
state of charge qubit is calculated. These analytic calculations show
how nonlinearity parameter and decay rates of two different excited
states of the qubit-resonator system affect time-oscillating and decaying
of Rabi oscillation.
\end{abstract}



\maketitle

\section{Introduction}

Actually, any quantum system interacting with the environment (the
bath) can not be isolated from the environment completely. \cite{Quantum Ddecoherence}
In quantum information and quantum computation, \cite{QInformation}
the decay process of quantum system induced by quantum fluctuations
of the bath is very important for the qubit. In quantum optics, \cite{Quantum Optics}
the Jaynes-Cummings model has been one of the most important models, \cite{JC model}
which describes the light-matter interaction of a two-level atom and
a single mode of the quantized electromagnetic field. \cite{Cavity QED}
Among these light-matter interaction issues, \cite{Atom-photon interaction}
the revivals and collapses of the atomic population inversion (also
named Rabi oscillation) has been studied in the literatures. \cite{Rabi-1,Rabi-2,Rabi-3-experment}

Decay of Rabi oscillation has also been used as a tool to characterize
the decoherence in superconducting qubits (charge qubit, phase qubit
and flux qubit). \cite{JJ qubit,JJ qubit-Nori} Recently, in circuit
QED system, \cite{Circuit QED} the researchers have performed spectroscopic
measurements of a superconducting qubit dispersively coupled to a
nonlinear resonator driven by a pump microwave field. \cite{Circuit QED with Nonlinear Resonator}
Also in nanomechanical QED system, the integration of Josephson junction
qubit and nanomechanical resonators are attracting considerable attentions. \cite{Nanomechanical QED,Nanomechanical Resonator,Nanomechanical Resonator approaching quantum limit,Roukes13}
The dynamics of all these qubit-resonator systems could be described
by the Jaynes-Cummings Hamiltonian.

When intrinsic nonlinearity of nanomechanical resonator \cite{Source of Nonlinearity}
is considered in the coupled qubit-resonator system, superconducting
qubit can be used to probe quantum fluctuations of nonlinear
resonator. \cite{Quantum Heating of a Nonlinear Resonator} And the nonlinearity can be used to create nonclassical states in
mechanical systems \cite{nonlinearity to creast nonclassical state, Fock state in mechanical freedom}
and selectively address the nanomechanical qubit transitions in quantum information processing. \cite{Qinformation with NR Qubit}

In previous studies, \cite{An Open System Approach to Quantum Optics}
master equation approach has been used to deal with the issues in open quantum
system. In this paper, considering the influence of the environment
on this nanomechanical QED system, we can use microscopic master equation
approach \cite{Microscopic master equation} to solve time evolution
of the density operator for the qubit-resonator system and study the
temporal Behavior of Rabi Oscillation.

The paper is structured as follows. In sec.II, a nonlinear Jaynes-Cummings
Model \cite{Nonlinear JC model} is used to describe the dynamics of
the coupled qubit-nanomechanical resonator system. In sec.III, using
microscopic master equation approach, we solve time evolution of density
operator for the qubit-resonator system. The probability on excited
state of the qubit is calculated to show the temporal process of Rabi
oscillation. Finally, the results are summarized.

\section{the qubit-resonator system}

In nanomechanical QED system, we can use a Jaynes-Cummings type Hamiltonian
to describes the dynamics of the qubit-resonator system consisting
of a charge qubit and a nanomechanical resonator system,
\begin{equation}
H_{JC}=\frac{\omega_{q}}{2}\sigma_{z}+g\left(a\sigma_{+}+a^{\dagger}\sigma_{-}\right)+\omega_{c}a^{\dagger}a.\label{JC Ham}
\end{equation}

Considering the nonlinearity of nanomechanical resonator, the Hamiltonian
for this qubit-resonator system writes \cite{Nonlinear Quantum Decoherence}
\begin{equation}
H_{S}=H_{JC}+\chi a^{\dagger}a+\chi\left(a^{\dagger}a\right)^{2}.\label{Nonlinear JC Ham}
\end{equation}
Here the rotating-wave approximation ($\omega_{q}=\omega_{a}=\omega$)
and $\hbar=1$ is adopted. Corresponding to charge qubit and nanomechanical
resonator, the lowering (raising) operator $\sigma_{-}$ ($\sigma_{+}$)
and the annihilation (creation) operator $a$ ($a^{\dagger}$) satisfy
the commutation relation $[\sigma_{-},\sigma_{+}]=\sigma_{z}$ and
$[a,a^{\dagger}]=1$. The Hamiltonian in Eq. (\ref{Nonlinear JC Ham})
describes the dynamics of a nonlinear Jaynes-Cummings model, \cite{Nonlinear JC model}
and a quartic potential $x^{4}$ \cite{Source of Nonlinearity} gives
nonlinear part $\chi\left(a^{\dagger}a\right)^{2}$ which leads to
the phonon-phonon interaction in nanomechanical QED systems. The $g$
is the coupling constant and the $\chi$ is the nonlinearity parameter
($\chi\ll g$).

Solving the Hamiltonian $H_{S}$, we get the ground state $\vert E_{0}\rangle=\vert00\rangle$
with energy $E_{0}=-\omega/2$ and excited state doublets
\begin{eqnarray*}
\left\vert E_{1+}\right\rangle  & = & \cos\left(\frac{\theta}{2}\right)\left\vert 10\right\rangle +\sin\left(\frac{\theta}{2}\right)\left\vert 01\right\rangle ,\\
\left\vert E_{1-}\right\rangle  & = & -\sin\left(\frac{\theta}{2}\right)\left\vert 10\right\rangle +\cos\left(\frac{\theta}{2}\right)\left\vert 01\right\rangle ,
\end{eqnarray*}
for $n=1,2,3,...$ with energy
\[
E_{1\pm}=\left(\frac{\omega}{2}+\chi\right)\pm\Omega.
\]
Some parameters are defined, i.e., $\Omega=\sqrt{g^{2}+\chi^{2}}$
and $\theta=\arcsin\left(g/\Omega\right)$.

With the loss of nanomechanical resonator, the total Hamiltonian

\[
H_{\text{total}}=H_{S}+H_{I}+H_{B}
\]
consists of three parts, i.e., the system part $H_{S}$, the interaction
part
\[
H_{I}=\sum_{j}\xi_{j}\left(ab_{j}^{\dagger}+a^{\dagger}b_{j}\right)
\]
and the bath part
\[
H_{B}=\sum_{j}\omega_{j}b_{j}^{\dagger}b_{j}.
\]
Where $b_{j}$ and $b_{j}^{\dagger}$ are bosonic annihilation and
creation operators for the bath oscillators for the mode frequency
$\omega_{j}$ ($j=1,2,...$).

In this paper, we adopt microscopic master equation approach \cite{Microscopic master equation}
to solve time evolution of density operator ($\rho$) for the qubit-resonator
system, our master equation is

\begin{equation}
\dot{\rho}=\mathcal{L}\rho\label{master equation-1}
\end{equation}
where $\mathcal{L}$ is a time-independent linear superoperator.

Using the microscopic master equation approach, \cite{Microscopic master equation}
we obtain the eigen-equations

\begin{equation}
\mathcal{L}\rho_{k}=\lambda_{k}\rho_{k}.\label{eigen operator}
\end{equation}
The $\left\{ \rho_{k}\right\} $ is a set of eigenoperators due to
the superoperator $\mathcal{L}$ with the eigenvalue $\left\{ \lambda_{k}\right\} $
for the index $k$.

Given initial state of the qubit-resonator system, the initial reduced
density operator $\rho\left(0\right)$ is expanded in terms of $\rho_{k}$,
\begin{equation}
\rho\left(0\right)=\sum_{k}C_{k}\rho_{k}\label{Ck}
\end{equation}
where the $C_{k}$s are time-independent coefficients. The results
in Ref.\cite{Microscopic master equation} tell us that time evolution
of reduced density operator $\rho$ will be

\begin{equation}
\rho\left(t\right)=\sum_{k}C_{k}e^{_{\lambda_{k}t}}\rho_{k}.\label{time rho}
\end{equation}

Now only one excitation is interested, our truncated basis $\left\{ \vert E_{1+}\rangle,\vert E_{1-}\rangle,\vert E_{0}\rangle\right\} $
consists of the three lowest eigenstates due to the Hamiltonian $H_{S}$,

Now we can rewrite the master equation in Eq. (\ref{master equation-1}),

\begin{equation}
\dot{\rho}=-i\left[H_{S},\rho\right]+\mathcal{L}_{+}\rho+\mathcal{L}_{-}\rho.\label{master equation}
\end{equation}
Here the non-unitary parts of dissipative dynamics are described by
$\mathcal{L}_{+}\rho$ and $\mathcal{L}_{-}\rho$,

\begin{eqnarray*}
\mathcal{L_{\pm}\rho} & = & \frac{\gamma_{\pm}}{2}\left\vert E_{0}\right\rangle \left\langle E_{1\pm}\right\vert \rho\left\vert E_{1\pm}\right\rangle \left\langle E_{0}\right\vert \\
 &  & -\frac{\gamma_{\pm}}{4}\left(\left\vert E_{1\pm}\right\rangle \left\langle E_{1\pm}\right\vert \rho+\rho\left\vert E_{1\pm}\right\rangle \left\langle E_{1\pm}\right\vert \right).
\end{eqnarray*}
The superoperator $\mathcal{L}_{\pm}$describe the transitions between
the higher excited state $\vert E_{\pm}\rangle$ and the ground state
$\vert E_{0}\rangle$ induced by the environment. The corresponding
decay rate $\gamma_{+}$ ($\gamma_{-}$) describes the transition
from the excited state $\vert E_{1+}\rangle$ ($\vert E_{1-}\rangle$)
to the ground state $\vert E_{0}\rangle$, these transitions are induced
by the interaction between the system and the environment.

With respect to the system Hamiltonian $H_{S}$ in Eq. (\ref{Nonlinear JC Ham}),
the eigenoperators $\rho_{k}$s are obtained,
\begin{eqnarray*}
\rho_{1} & = & \left\vert E_{0}\right\rangle \left\langle E_{0}\right\vert ,\\
\rho_{2} & = & \left\vert E_{1,-}\right\rangle \left\langle E_{1,-}\right\vert -\left\vert E_{0}\right\rangle \left\langle E_{0}\right\vert ,\\
\rho_{3} & = & \left\vert E_{1,+}\right\rangle \left\langle E_{1,+}\right\vert -\left\vert E_{0}\right\rangle \left\langle E_{0}\right\vert ,\\
\rho_{4} & = & \left\vert E_{0}\right\rangle \left\langle E_{1,-}\right\vert ,\\
\rho_{5} & = & \left\vert E_{0}\right\rangle \left\langle E_{1,+}\right\vert ,\\
\rho_{6} & = & \left\vert E_{1,-}\right\rangle \left\langle E_{1,+}\right\vert ,\\
\rho_{7} & = & \rho_{4}^{\dagger},\ \rho_{8}=\rho_{5}^{\dagger},\ \rho_{9}=\rho_{6}^{\dagger}.
\end{eqnarray*}
The corresponding eigenvalues $\lambda_{k}$s (for $k=1,2,3,...,9$)
are
\begin{eqnarray*}
\lambda_{1} & = & 0,\ \lambda_{2}=-\frac{\gamma_{1-}}{2},\ \lambda_{3}=-\frac{\gamma_{1+}}{2},
\end{eqnarray*}
\begin{eqnarray*}
\lambda_{4} & = & i\left(\omega+\chi-\Omega\right)-\frac{1}{4}\gamma_{1-},\\
\lambda_{5} & = & i\left(\omega+\chi+\Omega\right)-\frac{1}{4}\gamma_{1+},\\
\lambda_{6} & = & i(2\Omega)-\frac{1}{4}\left(\gamma_{1+}+\gamma_{1-}\right),
\end{eqnarray*}
and
\[
\lambda_{7}=\lambda_{4}^{*},\ \lambda_{8}=\lambda_{5}^{*},\ \lambda_{9}=\lambda_{6}^{*}.
\]

\section{Rabi Oscillation}

In traditional cavity QED theory, \cite{Cavity QED} the Rabi oscillation
means that there exists energy exchange of one photon between a two-level
atom and a single mode quantized field in cavity. Considering the
nonlinearity of nanomechanical resonator, we study the decay process
of Rabi oscillation in the nonlinear Jaynes-Cummings model described
by the Hamiltonian in Eq. (\ref{Nonlinear JC Ham}).

Given the initial state of the qubit-resonator system $\vert\psi\left(0\right)\rangle=\vert e\rangle\otimes\vert0\rangle$,
it means that the qubit is in excited state $\vert e\rangle$ and
the resonator is in vacuum state $\vert0\rangle$. Thus, the initial
reduced density operator reads
\[
\rho\left(0\right)=\left\vert \psi\left(0\right)\right\rangle \left\langle \psi\left(0\right)\right\vert .
\]
Expanding $\rho\left(0\right)$ into some eigenoperators $\rho_{k}$s,
we obtain a set of coefficients $C_{k}$s,
\[
C_{1}=1,\ C_{2}=\frac{1}{2}\left(1-\cos\theta\right),\ C_{3}=\frac{1}{2}\left(1+\cos\theta\right),
\]
\[
C_{6}=C_{9}=-\frac{1}{2}\sin\theta,
\]
and
\[
C_{4}=C_{5}=C_{7}=C_{8}=0.
\]

According to Eq.(\ref{time rho}), the time evolution of density operator
for the qubit-resonator system is calculated as

\begin{eqnarray*}
\rho\left(t\right) & = & C_{1}e^{_{\lambda_{1}t}}\rho_{1}+C_{2}e^{_{\lambda_{2}t}}\rho_{2}\\
 &  & +C_{3}e^{_{\lambda_{3}t}}\rho_{3}+C_{6}e^{_{\lambda_{6}t}}\rho_{6}+C_{9}e^{_{\lambda_{9}t}}\rho_{9}.
\end{eqnarray*}
Here the probability of the qubit in the excited (upper) state $\left\vert e\right\rangle $
is

\begin{eqnarray}
P_{e}(t) & = & \left\langle e0\right\vert \rho\left(t\right)\left\vert e0\right\rangle \nonumber \\
 & = & \left[\frac{1}{2}\left(1-\cos\theta\right)e^{-\frac{\gamma_{1-}}{4}t}-\frac{1}{2}\left(1+\cos\theta\right)e^{-\frac{\gamma_{1+}}{4}t}\right]^{2}\nonumber \\
 &  & +\sin^{2}\theta e^{-\frac{\gamma_{1+}+\gamma_{1-}}{4}t}\cos^{2}\left(\Omega t\right).\label{probability of excited state}
\end{eqnarray}
It characterizes the temporal behavior of Rabi oscillation in the
qubit-resonator system, decay process of Rabi oscillation owns the
periodic structure of time oscillating.

The nanomechanical resonator is assumed to be an ideal resonator ($\chi=0$),
and ignoring the difference of decay rates ($\gamma_{1+}=\gamma_{1-}=\gamma$),
then the probability $P_{e}(t)$ becomes

\begin{equation}
P_{e}(t)=e^{-\frac{\gamma}{2}t}\cos^{2}\left(gt\right).\label{probability at excited state in Rabi}
\end{equation}
It describes the well known Rabi oscillation in Jaynes-Cummings model.\cite{JC model}
Comparing the results in Eq. (\ref{probability of excited state})
with Eq. (\ref{probability at excited state in Rabi}), we find that
nonlinearity parameter $\chi$ and decay rates $\gamma_{1+}\neq\gamma_{1-}$
modify the periodic structure of time oscillating in Rabi oscillation.

To further clarify the dependence of nonlinearity parameter and decay
rates on the probability $P_{e}(t)$ clearly, some figures are plotted
with parameters $\omega=1.0,\ g=0.1$. Here, we take the frequency
$\omega$ as the unit for all these parameters.

In Fig.$\,1$, the probability $P_{e}(t)$ versus time $t$ is plotted
with parameters $\chi=0$ and $\gamma_{1+}=\gamma_{1-}=0.004$. Figure
1 shows the well known Rabi oscillation, it verifies the results in
Eq. (\ref{probability at excited state in Rabi}).

\begin{figure}[ht]
\includegraphics[width=8cm]{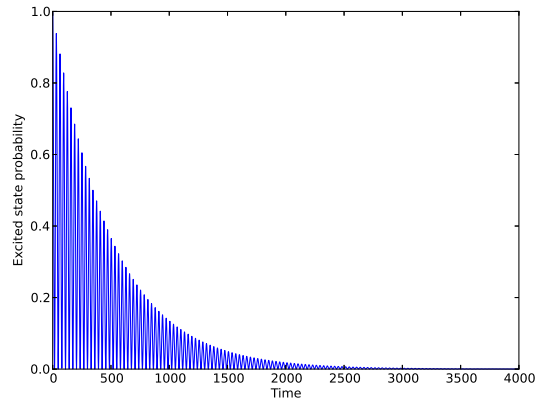}\label{Rabi in JC model} \caption{Rabi oscillation in Jaynes-Cummings model. Some parameters are $\chi=0$
and $\gamma=0.004$.}
\end{figure}

\begin{figure}[ht]
\includegraphics[width=8cm]{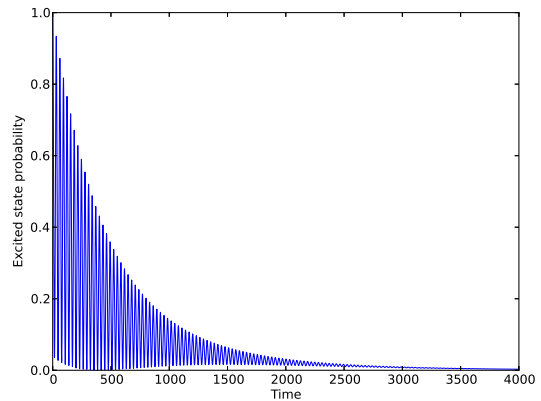}\label{different decay rate and nonlinearity}
\caption{The probability $P_{e}\left(t\right)$ vs the time $t$, where the
excite states ($\vert1+\rangle$ and $\vert1-\rangle$) own the different
decay rate ($\gamma_{1-}=0.001$ and $\gamma_{1+}=0.007$) and $\chi=0$.04.}
\end{figure}

In Fig.$\,2$, the probability $P_{e}(t)$ versus time $t$ is plotted
with parameters $\chi=0.04$, $\gamma_{1+}=0.007$ and $\gamma_{1-}=0.001$.
Figure 2 shows that nonlinearity parameter $\chi$ and decay rates
($\gamma_{1+}\neq\gamma_{1-}$) modify the periodic structure of time
oscillating in Rabi oscillation, which is obviously different from
Figure 1.

Based on those results in Eq. (\ref{probability of excited state},\ref{probability at excited state in Rabi}),
we find that nonlinearity parameter $\chi$ slows down the time-oscillating
period of Rabi oscillation $T=\left(2\Omega\right)^{-1}$. In the
following, we will study how nonlinearity parameter $\chi$ and decay
rates ($\gamma_{1+}\neq\gamma_{1-}$) affect the temporal behavior
in Rabi oscillation solely.

Firstly, assuming the same decay rates $\gamma_{1-}=\gamma_{1+}=\gamma$
and nonlinearity parameter $\chi\neq0$, the probability in Eq. (\ref{probability of excited state})
becomes
\begin{equation}
P_{e}(t)=e^{-\frac{\gamma}{2}t}\left(\cos^{2}\theta+\sin^{2}\theta\cos^{2}\left(\Omega t\right)\right).\label{Pe of same decay rate}
\end{equation}
The dependence of the probability $P_{e}(t)$ on nonlinearity parameter
$\chi$ is plotted in Fig.$\,3$. When $\cos^{2}\left(\Omega t\right)=0$,
the minimum of the probability
\begin{equation}
P_{e}(t)=e^{-\frac{\gamma}{2}t}\cos^{2}\theta\label{Pe of same decay rate-1}
\end{equation}
decays exponentially, which is different from the well known Rabi
oscillation in Fig.$\,1$.

Secondly, assuming no nonlinearity $\chi=0$ and different decay rates
($\gamma_{1+}\neq\gamma_{1-}$), the probability in Eq. (\ref{probability of excited state})
becomes
\[
P_{e}(t)=\frac{1}{4}\left(e^{-\frac{\gamma_{1-}}{4}t}-e^{-\frac{\gamma_{1+}}{4}t}\right)^{2}+e^{-\frac{\gamma_{1+}+\gamma_{1-}}{4}t}\cos^{2}\left(gt\right).
\]
The dependence of the probability $P_{e}(t)$ on different decay rates
is plotted in Fig.$\,4$. When $\cos^{2}\left(gt\right)=0$, the minimum
of the probability
\[
P_{e}(t)=\frac{1}{4}\left(e^{-\frac{\gamma_{1-}}{4}t}-e^{-\frac{\gamma_{1+}}{4}t}\right)^{2}
\]
shows that the difference of decay rates between $\gamma_{1-}$ and
$\gamma_{1+}$ does not affect the short-time behavior and long-time
behavior of Rabi oscillation obviously, which is seen in Fig.$\,4$.

\begin{figure}[ht]
\includegraphics[width=8cm]{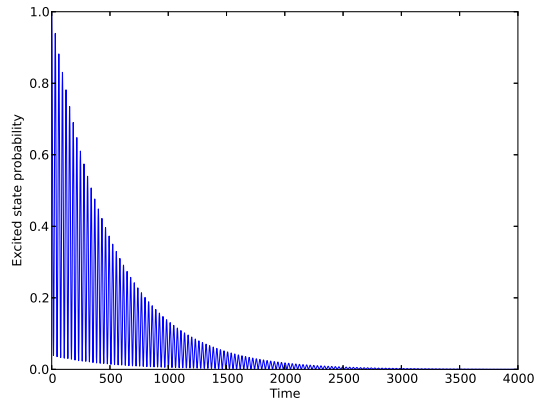}\label{nonlinearity fig1}
\caption{The probability $P_{e}\left(t\right)$ vs the time $t$, where the
excite states ($\vert1+\rangle$ and $\vert1-\rangle$) own the same
decay rate ($\gamma_{1+}=\gamma_{1-}=\gamma=0.004$) and nonlinearity
parameter $\chi=0.04$.}
\end{figure}

\begin{figure}[ht]
\includegraphics[width=8cm]{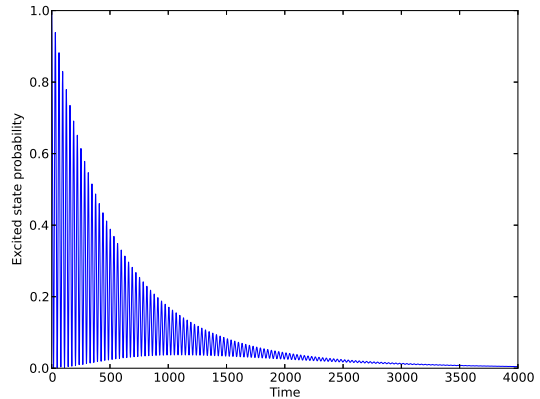}\label{pure JC with microscopic master}
\caption{The probability $P_{e}\left(t\right)$ vs the time $t$, the excite
states ($\vert1+\rangle$ and $\vert1-\rangle$) of the qubit-resonator
system own the different decay rate ($\gamma_{1-}=0.001$ and $\gamma_{1+}=0.007$)
and nonlinearity parameter $\chi=0$.}
\end{figure}

Analytically, we can study the short-time behavior of Rabi oscillation,
then the probability $P_{e}\left(t\right)$ becomes
\begin{equation}
P_{e}(t)=\exp\left\{ -\left[\cos\theta\left(\frac{\gamma_{1+}}{4}-\frac{\gamma_{1-}}{4}\right)+\frac{\gamma_{1-}+\gamma_{1+}}{4}\right]t\right\} .\label{short-time probability}
\end{equation}
Ignoring the nonlinearity of nanomechanical resonator or the difference
of decay rates, i.e., $\chi=0$ or $\gamma_{1-}=\gamma_{1+}$, the
probability $P_{e}\left(t\right)$ becomes
\begin{equation}
P_{e}(t)=\exp\left\{ -\frac{\gamma_{1-}+\gamma_{1+}}{4}t\right\} .\label{short-time decay rate}
\end{equation}
According to the results in Eq. (\ref{short-time probability}) and
Eq. (\ref{short-time decay rate}), we find that both nonlinearity
parameter $\chi$ and the difference of decay rates ($\gamma_{1+}\neq\gamma_{1-}$)
affect dominate the short-time behavior of Rabi oscillation jointly.
Also, these two factors speed up the decay of Rabi oscillation in
short-time limit.

\section{conclusions}

In summary, we have studied the dynamics of the nanomechanical QED
system consisting of a charge qubit and a nanomechanical resonator.
The temporal behavior of Rabi oscillations is analytically studied
while the intrinsic nonlinearity of nanomechanical resonator is considered.
With the loss of nanomechanical resonator, microscopic master equation
approach is used to calculate the excited-state probability of charge
qubit in this nonlinear Jaynes-Cummings model. These results show
that nonlinearity parameter and decay rates can affect time-oscillating
and decaying of Rabi oscillation solely or jointly.

\begin{acknowledgments}
We thank the discussions with Professor Peng Zhang and Dr. Ming Hua.
\end{acknowledgments}


\begin{thebibliography}{DiVincenzo(2000)}
\bibitem[Zurek(1983)]{Quantum Ddecoherence} J. A. Wheeler and Z.
H. Zurek, \emph{Quantum Theory of Measurement}, Princeton University
Press, NJ (1983).

\bibitem[DiVincenzo(2000)]{QInformation} D. DiVincenzo, Fortschr.
Phys. \textbf{48} (2000) 771.

\bibitem[Scully(1997)]{Quantum Optics} M. O. Scully and M. S. Zubariry,
Quantum Optics, Cambridge University Press, Cambridge, 1997.

\bibitem[Jaynes(1963)]{JC model} E. T. Jaynes and F. W. Cummings,
Proc. IEEE \textbf{51} (1963) 89.

\bibitem[Raimond(2001)]{Cavity QED} J. M. Raimond, M. Brune, and
S. Haroche, Rev. Mod. Phys. \textbf{73} (2001) 565.

\bibitem[Cohen(1998)]{Atom-photon interaction} C. Cohen-Tannoudji,
J. Dupont-Roc, and G. Grynberg, \textit{Atom-Photon Interactions}
(John Wiley, New York, 1998).

\bibitem[Eberly(1980)]{Rabi-1} J. H. Eberly, N. B. Narozny, and J.
J. Sanchez- Mondragon, Phys. Rev. Lett.\textbf{ 44} (1980) 1323.

\bibitem[Eberly(1981)]{Rabi-2} N. B. Narozny, J. J. Sanchez-Mondragon,
and J. H. Eberly, Phys. Rev. A. \textbf{23} (1981) 236.

\bibitem[Rempe(1987)]{Rabi-3-experment} G. Rempe, H. Walther, and
N. Klein, Phys. Rev. Lett. \textbf{58} (1987) 353.

\bibitem[Makhlin(2001)]{JJ qubit} Y. Makhlin, G. Schoen, and A. Shnirman,
Rev. Mod. Phys. \textbf{73} (2001) 357.

\bibitem[Nori(2005)]{JJ qubit-Nori} J. Q. You and F. Nori, Phys.
Today \textbf{58} (11) (2005) 42.

\bibitem[Blais(2004)]{Circuit QED} A. Wallraff, D. I. Schuster, A.
Blais, L. Frunzio, R. S. Huang, J. Majer, S. Kumar, S. M. Girvin,
and R. J. Schoelkopf, Nature \textbf{431} (2004) 162.

\bibitem[Ong(2011)]{Circuit QED with Nonlinear Resonator} F. R. Ong,
M. Boissonneault, F. Mallet, A. Palacios-Laloy, A. Dewes, A. C. Doherty,
A. Blais, P. Bertet, D. Vion, and D. Esteve, Phys. Rev. Lett. \textbf{106}
(2011) 167002.

\bibitem[Gao(2009)]{Nanomechanical QED} Y. B. Gao, S. Yang, Y. X.
Liu, C. P. Sun, and F. Nori, arxiv:0902.2512.

\bibitem[Xue(2007)]{Nanomechanical Resonator} F. Xue, Y. D. Wang,
C. P. Sun, H. Okamoto, H. Yamaguchi, and K. Semba, New J. Phy. \textbf{9}
(2007) 35.

\bibitem[Roukes(2009)]{Nanomechanical Resonator approaching quantum limit}
M. D. LaHaye, J. Suh, P. M. Echternach, K. C. Schwab, and M. L. Roukes,
Nature \textbf{459} (2009) 960.

\bibitem[Roukes(2013)]{Roukes13} L. G. Villanueva, R. B. Karabalin,
M. H. Matheny, D. Chi, J. E. Sader, and M. L. Roukes, Phys. Rev. B
\textbf{87} (2013) 024304.


\bibitem[Hartmann(2014)]{Source of Nonlinearity} S. Rips, I. WilsonRae,
and M. J. Hartmann, Phys. Rev. A \textbf{89} (2014) 013854.

\bibitem[Blais(2013)]{Quantum Heating of a Nonlinear Resonator} F.
R. Ong, M. Boissonneault, F. Mallet, A. C. Doherty, A. Blais, D. Vion,
D. Esteve, and P. Bertet, Phys. Rev. Lett. \textbf{110} (2013) 047001.

\bibitem[Hartmann(2015)]{Fock state in mechanical freedom} M. Abdi, M. Pernpeintner, R. Gross, H. Huebl, and M. J. Hartmann Phys. Rev. Lett. \textbf{114} (2015) 173602.

\bibitem[Paris(2015)]{nonlinearity to creast nonclassical state}
B. Teklu, A. Ferraro, M. Paternostro, and M. G. A. Paris, arXiv:1501.03767.

\bibitem[Hartmann(2013)]{Qinformation with NR Qubit} S. Rips and M. J. Hartmann, Phys. Rev. Lett. \textbf{111} (2013) 049905.


\bibitem[Camichael(1993)]{An Open System Approach to Quantum Optics}
H. J. Carmichael, \textit{Lecture Notes in Physics Springer-Verlag},
Berlin, Heidelberg, 1993.

\bibitem[Scala(2007)]{Microscopic master equation} M. Scala, B. Militello,
A. Messina, J. Piilo, S. Maniscalco, Phys. Rev. A \textbf{75} (2007)
013811.

\bibitem[P Gora(1992)]{Nonlinear JC model} P. G\'{o}ra and C. Jedrzejek,
Phys. Rev. A \textbf{45} (1992) 6816.

\bibitem[Gao(2013)]{Nonlinear Quantum Decoherence} C. Chen and Y.
B. Gao, Commun. Theor. Phys. \textbf{60} (2013) 531.

\end{thebibliography}
\end{document}